\documentclass[11pt,twoside]{article}

\usepackage{asp2006}
\usepackage{natbib}
\usepackage{epsf}
\usepackage{lscape}
\usepackage{graphicx}

\begin{document}
\title{The New Solar Minimum: How deep does the problem go}   %%% Fill in title
\author{Stephen Fletcher and Roger New}   %%% Fill in author names
\affil{Sheffield Hallam University, Sheffield, UK}    %%% Fill in author affiliations
%\affil{Faculty of Arts, Computing, Engineering and Science,
%Sheffield Hallam University, Sheffield, UK}    %%% Fill in author affiliations
\author{Anne-Marie Broomhall, William Chaplin and Yvonne Elsworth}   %%% Fill in author names
\affil{University of Birmingham, Birmingham, UK}    %%% Fill in author affiliations
%\affil{School of Physics and Astronomy, University of Birmingham,
%Birmingham, UK}    %%% Fill in author affiliations

\begin{abstract} %%% Abstract to run on from here.
Although there are now some tentative signs that the start of cycle
24 has begun there is still considerable interest in the somewhat
unusual behaviour of the current solar minimum and the apparent
delay in the true start of the next cycle. While this behaviour is
easily tracked by observing the change in surface activity a
question can also be asked about what is happening beneath the
surface where the magnetic activity ultimately originates? In order
to try to answer this question we can look at the behaviour of the
frequencies of the Sun's natural seismic modes of oscillation - the
p modes. These seismic frequencies also respond to changes in
activity and are probes of conditions in the solar interior.

The Birmingham Solar Oscillations Network (BiSON) has made
measurements of low-degree (low-$\ell$) p mode frequencies over the
last three solar cycles, and so is in a unique position to explore
the current unusual and extended solar minimum. We compare the
frequency shifts in the low-$\ell$ p-modes obtained from the BiSON
data with the change in surface activity as measured by different
proxies and show there are significant differences especially during
the declining phase of solar cycle 23 and into the current minimum.
We also observe quasi-biennial periodic behaviour in the p mode
frequencies over the last 2 cycles that, unlike in the surface
measurements, seems to be present at mid- and low-activity levels.
Additionally we look at the frequency shifts of individual $\ell$
modes.
\end{abstract}

\section{Introduction}

The magnetic activity of the Sun is known to follow an approximately
11-year cycle ranging from quiet periods to much more active
periods. Starting in March 1755 these cycles have been labeled with
a number and we are currently in the minimum between cycles 23 and
24. However, this current minimum is proving to be somewhat unusual,
both in terms of how long it is lasting and just how quiet the Sun
is, at least in terms of the more recently observed cycles.

Surface measures of solar actively such as the number of visible
sunspots and 10.7cm radio flux, all suggest that we are still in (or
possibly just leaving) an extended minimum. Therefore, the question
presents itself is there anything different about this minimum that
we can learn from looking at the activity of the solar interior.

This can be investigated by looking at the acoustic modes of
oscillation of the Sun (known as p modes). Modes with low angular
degrees (low-$\ell$) travel deeply into the solar interior and so
sample the conditions below the solar surface. Also, low-$\ell$
modes are truly global in nature and therefore are sensitive to a
large fraction of the solar disc. It has been known for some time
that the frequencies of these oscillations vary during the solar
cycle, with the frequencies being highest when the magnetic activity
is at a maximum. Therefore, by looking at the change in frequencies
we can gain insight into cycle related processes that are occurring
beneath the solar surface.

The Birmingham Solar Oscillations Network (BiSON) has been
collecting unresolved (Sun-as-a-star) Doppler velocity observations
for over three decades. These types of observations are sensitive to
the large scale `global' oscillations of the sun, (i.e., those modes
with the lowest angular degrees, $\ell$). Due to the fact the BiSON
has observations going back into the 1970's, these data offer a
unique opportunity to study the last three solar cycles. However,
when the network was first established there was only one station
and as such the early observations are sporadic and so the data
quality is relatively poor. The quality and duty cycle of the data
greatly improves after 1986 once three stations were operating and
this allows us to study the last two solar cycles in their entirety.

\section{Method}

The precision with which one can measure the p-mode frequencies is
directly related to the length of the time series. Therefore, one
often wants to take as long a time series as possible in order to
give the most precise estimates of the frequencies. However, when
trying to track the change in frequencies over the solar cycle, one
must strike a balance between length of time series and the time
resolution of each frequency measurement. To this end an 8577-day
BiSON time series starting on 14 April 1986 and ending on 7th
October 2009 was split into a set of 365-day time series. Each
series was allowed to overlap the following by 91.25 days (i.e., a
4-time overlap), resulting in a total of 91 (non-independent) data
sets. The power spectrum of each time series was parameterised using
a modified Lorentzian model which was fitted to the data using a
standard likelihood maximisation method \citep[eg.,
see][]{Chaplin1999} enabling the mode frequencies to be determined.

For each mode a minimum activity reference was taken from the fits
to a data set lying on the boundary between cycle 22 and 23. The
frequency shifts were then defined as the difference between this
reference value and the frequency of the corresponding mode observed
at different epochs. The size of the frequency shift has a known
dependence on frequency and this dependence was removed in the
manner described in \cite{Chaplin2007}. A weighted average of the
frequency shift for each time series was then determined, enabling
the shift to be plotted over time. The average was made over modes
ranging from $12 \leq n \leq 25$ and $0 \leq \ell \leq 3$.

A similar analysis was performed recently with BiSON data
\citep[see][]{Broomhall2009}. Here we extend in time the results of
that paper and additionally present the shifts of individual $\ell$
modes. This allows us to make comments about the location of the
solar variability, since each $\ell$ has different latitudinal
sensitivities.

\section{Analysis and Results}

\begin{figure}
\centerline{\includegraphics[width=4.5in]{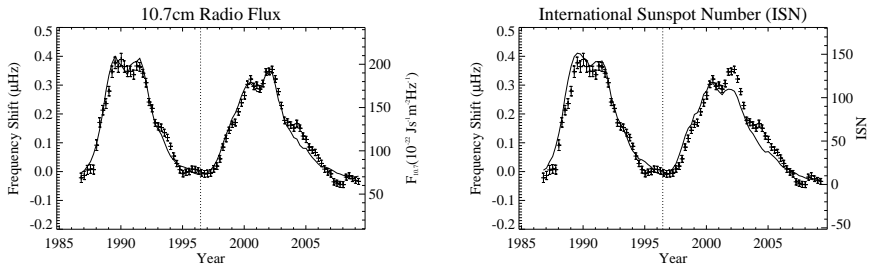}}\caption{Compariosn
between the frequency shifts as determined from BiSON data and two
different surface activity proxies. Left: The 10.7cm radio flux.
Right: The International Sunspot number, both shown by the solid
black line.} \label{lav}
\end{figure}

In this section we concentrate on the last two cycles (22 and 23)
where the BiSON data are of sufficient quality to do meaningful
statistical tests. It is widely known \citep[see][and references
therein]{Elsworth1990,Howe2008} that there is a good correlation
between the shifts in the p-mode frequencies and other proxies that
can be used to track the solar cycle. In Fig.~\ref{lav} we compare
the shifts with two of the more commonly used proxies, the 10.7cm
radio flux ($F_{10.7}$) and the International Sunspot Number (ISN).
The shifts show very good agreement with the 10.7cm radio flux up
until around 2003, after which there is a significant disagreement
relative to previous epochs. The agreement between the shifts and
the ISN is not as good. This is thought to be because the ISN is
predominantly sensitive to the strong component of the Sun's
magnetic field, whereas the frequency shifts and radio flux are
sensitive to both the strong \textit{and} weak components. However,
even with the poorer fit, it is still clear that the largest
differences still occur over the latter part of cycle 23.

Looking more closely at the final few points shown in
Fig.~\ref{lav}, it appears that there is a significant rise in the
frequency shifts at the start of 2008, that may be pointing to the
start of cycle 24. We do not see a similar rise in either the radio
flux or the ISN. However, this rise in the shifts did not continue,
and looking back at the boundary of the last cycle we can see there
are similar rises and dips in the shifts throughout the last minimum
too.

If these dips and rises in the shifts are not associated with the main
solar cycle then they do highlight an interesting notion in their
own right. There are clearly bumps and dips at higher activity in
both the shifts and other proxies, but if these continue into low
activity periods too, then it might be a an indication of short-term
pseudo periodic behaviour.

\begin{figure}
\centerline{\includegraphics[width=4.5in]{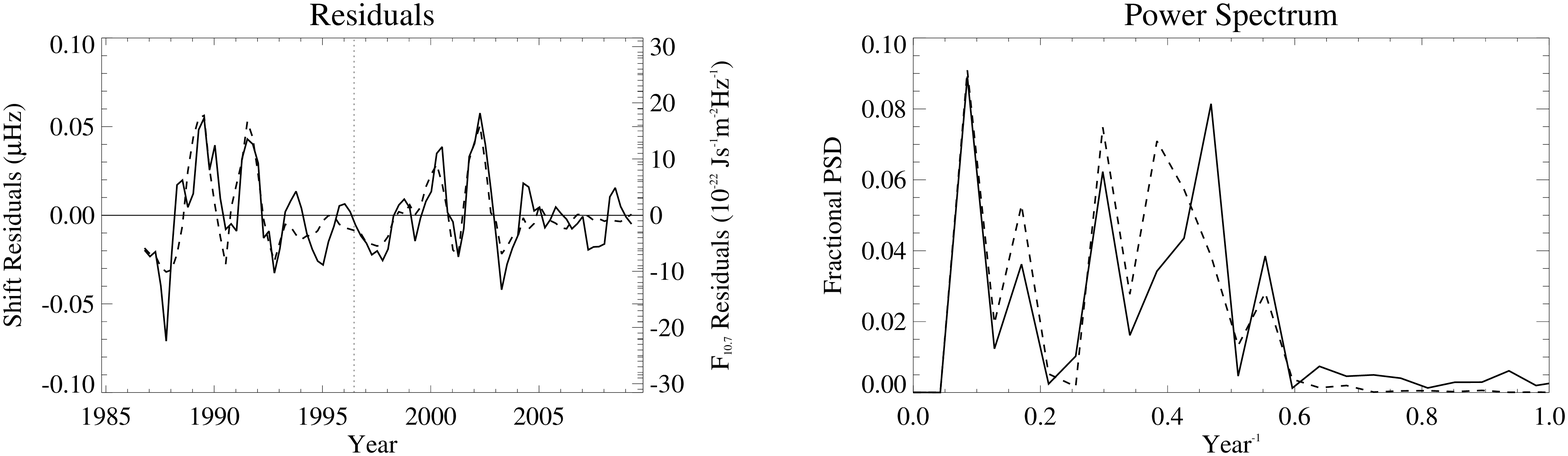}}\caption{Left:
The residuals obtained when subtracting the unsmoothed frequency
shifts (solid line) and radio flux (dotted line) from heavily
smoothed values. Right: The power spectrum of these residuals.}
\label{resid}
\end{figure}

To investigate this more thoroughly we have looked at the difference
between heavily smoothed and unsmoothed data. The residuals of this
comparison are shown in the left panel of Fig.~\ref{resid}. Short
term periodic variations appear to occur on a scale of around two
years. The variations seem to be stronger in the oscillations during
the minimum periods but of similar magnitude to the radio flux
during high activity. We can also look at a power spectrum of these
residuals as shown in the right panel of Fig.~\ref{resid}. Although
there is not a clear and obvious peak at short time periods there
are a cluster of peaks around a two to three year period. Similar
``quasi-biennial'' variation have been remarked upon before,
specifically in the green coronal emission line at 530.3nm at high
solar activity \citep{Vecchio2008}. Additionally, an explanation of
such quasi-biennial behaviour has been put forward by two different
types of dynamos operating at different depths
\citep{Benevolonskaya1998a,Benevolonskaya1998b}.

\begin{figure}
\centerline{\includegraphics[width=4.5in]{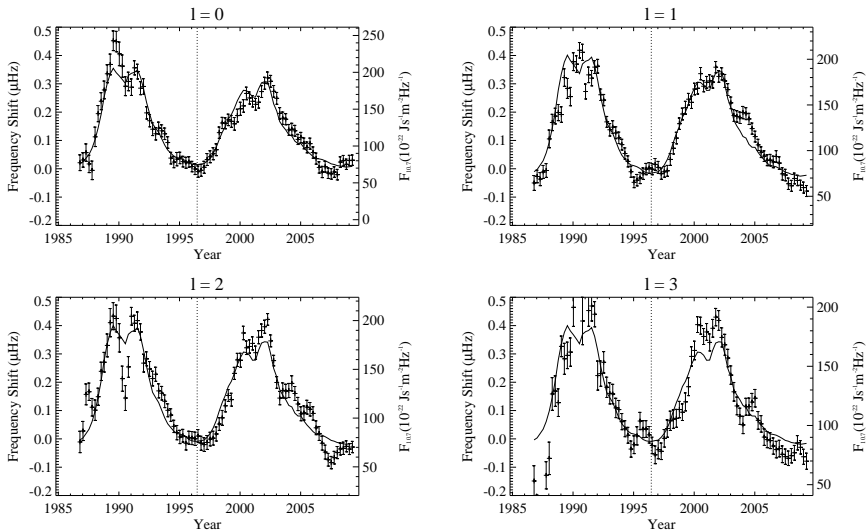}}\caption{Compariosn
between the frequency shifts as determined from BiSON data and the
10.7cm radio flux for each individual $\ell$-mode in the range $0
\leq \ell \leq 3$} \label{ldiff}
\end{figure}

We can gain further insight into what the shift in the frequencies
are telling us by looking at the shifts of individual angular
degrees. We show this for $\ell$ = 0 - 3 in Fig.~\ref{ldiff}. The
$\ell$ = 0 shifts seem to fit the radio flux very well, with a
reduced $\chi^2$ value of only 1.8 (which is considerably lower than
the averaged $\ell$ value of 3.5). This may be partly due to the
fact that the $\ell$ = 0 modes are more sensitive to the whole disc
of the Sun than the higher $\ell$ modes.

For the $\ell$ = 1 modes there is more structure than in the $\ell$
= 0 case and this results in a poorer fit with the radio flux (the
reduced $\chi^2$ is 3.3). The $\ell$ = 2 mode shifts show more
similarity with $\ell$ = 0 than $\ell$ = 1. However the fit to the
radio flux is again poorer (a reduced $\chi^2$ of 3.1). This
correlation between $\ell$ = 0 and $\ell$ = 2 modes and also $\ell$
= 1 and $\ell$ = 3 modes is an intriguing result. It might be due to
the fact that correlated modes have somewhat similar latitudinal
sensitivities. Both $\ell$ = 0 and $\ell$ = 2 modes have an $m$ = 0
zonal component which is sensitive to higher latitudes, whereas
$\ell$ = 1 and 3 modes only have sectoral components. Or
alternatively, the effect could be an artifact of the fitting
procedure since $\ell$ = 0 and 2 modes and $\ell$ = 1 and 3 modes
are fitted together in pairs. These ideas will be investigated
further in the future.

Again, we can look more closely at the final few points in each of
the plots in order to investigate how the current solar minimum and
start of cycle 24 is developing. For $\ell$ = 0 we see a clear jump
in the shifts at the beginning of 2008, but there is no obvious
decline after this. For $\ell$ = 1 the rise at the start of 2008 is
much less pronounced than for $\ell$ = 0, and there is a definite
decline after this. The $\ell$ = 2 shifts appear to start rising
before the beginning of 2008 and continue upwards thereafter.
Finally, for $\ell$ = 3 we see a short rise during 2008, although
the error bars are sufficiently large that the rises and dips are
not significant.

These results match quite closely with what is seen in an analysis
of GOLF data \citep[see][]{Salabert2009}. However, in the GOLF data
there was a clearer rise in the $\ell$ = 0 and 2 modes and less of a
dip in the $\ell$ = 1 data ($\ell$ = 3 modes were not included in
this analysis). \cite{Salabert2009} gave an explanation for this
based on the latitudinal sensitivity of the modes. They suggest
that, if it is assumed that at the time of the final few
measurements the start of cycle 24 had begun, then we would expect
any increased activity to occur at higher latitudes first. The
$\ell$ = 0 and $\ell$ = 2 modes are more sensitive to these higher
latitudes and hence may be seeing the effects of an increased
activity due to the onset of cycle 24. The $\ell$ = 1 modes, on the
other hand, whose sensitivities are more confined towards the
equator may not be seeing this increased activity yet.

The main argument against this explanation is that the zonal $m$ = 0
components of the $\ell$ = 2 multiplet, which will be the component
most sensitive to higher latitudes, only contributes about 10\% to
the determination of the fitted frequency. The frequency fit is
actually much more dependent on the sectoral $|m|$ = 2 components,
whose sensitivities are actually more concentrated around the
equator than the $\ell$ = 1 sectoral components. Therefore it would
seem that there is more investigation needed into the shifts of the
individual $\ell$ modes to try and understand these differences.
Also, as described above, we will investigate the pair-by-pair
nature of the fitting process to see whether it might be responsible
for our observations.

\section{Summary and Discussion}

The shifts in oscillation frequencies over time is reasonably well
correlated with other activity proxies such as the 10.7cm radio flux
and ISN. However, there are some differences. The largest of which,
at least over the two cycles we have data for, occur recently,
starting on the downward part of cycle 23 and heading into the
unusual and extended minimum between cycle 23 and 24. The current
minimum in the frequency shifts is considerably deeper than in the
proxies when compared with the previous minimum and there is clear
structure in the shifts that does not appear in the proxies.
Additionally, there is some evidence for pseudo-periodic short term
(2-year) variations in activity on top of the 11-year cycle.
Frequency shifts show sharper amplitude variations during the quiet
sun periods compared with the radio flux.

Since the frequency shifts are sensitive to conditions below the
surface it is likely that these differences are due to changes in
the magnetic flux that are occurring in the solar interior. These
changes then may have either yet to manifest on the surface, or are
attenuated before ever reaching there.

Alternatively the differences could be due to zonal effects, since
the oscillations are not all equally sensitive at different
latitudes of the Sun. We have looked at the shifts of modes with
different $\ell$ in order to gain a clearer picture of this. The
$\ell$ = 0 modes show the closest match with the radio flux. The
higher degree modes all show greater discrepancies. The strongest
regions of magnetic flux tend to start at latitudes near the poles
and then drift towards the equator as the cycle progresses.
Therefore, modes that are only sensitive to ceratin latitudes will
see their frequency shift's respond to this drift in magnetic flux.
This will ultimately add further structure into the plots of the
frequency shifts that may not be apparent in the $\ell$ = 0 modes or
radio flux plots.

If we assume that cycle 24 had started at the time of the last few
points in the data set used here, then the shape of the plots for
$\ell$ = 0, 1, 2, and 3 might also be explained by this latitudinal
sensitivity argument as suggested by \cite{Salabert2009}. Although
the fact that the $\ell$ = 2 modes showed a similar upturn to the
$\ell$ = 0 modes in the final few observations, may actually work
against this theory since the zonal $m$ = 0 component, which does
have a greater sensitivity to higher latitudes, only has a small
effect on the fitted frequency of the $\ell$ = 2 multiplet.

It is clear that further work on the individual $\ell$ modes is
needed to help clear up this matter. In the future we hope to
continue the work on individual $\ell$ modes by comparing the
frequency shifts with decomposed magnetic maps. This will allow us
to compare the frequency shifts with the change in magnetic flux
over time for similar zonal regions.

\acknowledgements %%% Text of acknowledgements runs on after this command.
STF acknowledges the support of the Faculty of Arts, Computing,
Engineering and Science (ACES) at Sheffield Hallam University and
the support of the Science and Technology Facilities Council (STFC).
We also thank all those associated with BiSON which is funded by the
STFC. The authors would also like to thank D. Salabert and R. A.
Garc{\'{\i}}a for sharing their insights and thoughts into the
analysis of GOLF data.

%%% THE BIBLIOGRAPHY
%%%
%%% CONSULT SECTION 3 OF "INSTRUCTIONS FOR AUTHORS" FOR HOW TO USE NATBIB.
%%% AUTHORS ARE ENCOURAGED TO USE EITHER THE "THEBIBLIOGRAPY" ENVIRONMENT
%%% BY UNCOMMENTING (DELETING THE "%" SYMBOL) THE COMMANDS BELOW, OR BY
%%% USING THE BIBTEX ENVIRONMENT. TO FIND OUT WHICH IS APPLICABLE TO YOUR
%%% CONTRIBUTION, CONSULT THE VOLUME EDITORS FOR YOUR PROCEEDINGS.
%%%

%\bibliography{fletcher}

\end{document}